\documentclass[superscriptaddress,pre,twocolumn,showpacs,showkeys]{revtex4}

\usepackage{amscd}

\newcommand{\hmu}{h_{\mu}}
\newcommand{\Cmu}{C_{\mu}}
\newcommand{\Prob}{\mathrm{Pr}}

\begin{document}
\title{Symbolic Dynamics for Discrete Adaptive Games}
\author{Cosma Rohilla Shalizi}
\email{cshalizi@umich.edu}
\affiliation{Center for the Study of Complex Systems, University of
Michigan, Ann Arbor, MI 48109}
\affiliation{Santa Fe Institute, 1399 Hyde Park Road, Santa Fe, NM 87501}
\author{David J. Albers}
\email{albers@cow.physics.wisc.edu}
\affiliation{Physics Department, University of Wisconsin, Madison, WI 53706}
\affiliation{Santa Fe Institute, 1399 Hyde Park Road, Santa Fe, NM 87501}
\date{\today}

\begin{abstract}
We use symbolic dynamics to study discrete adaptive games, such as the minority
game and the El Farol Bar problem.  We show that no such game can have
deterministic chaos.  We put upper bounds on the statistical complexity and
period of these games; the former is at most linear in the number agents and
the size of their memories.  We extend our results to cases where the players
have infinite-duration memory (they are still non-chaotic) and to cases where
there is ``noise'' in the play (leaving the complexity unchanged or even
reduced).  We conclude with a mechanism that can reconcile our findings with
the phenomenology, and reflections on the merits of simple models of mutual
adaptation.
\end{abstract}



\maketitle

\section{Discrete Adaptive Games}

\subsection{Introduction}

In the last few years, physicists have become very interested in discrete
games.  In these models, a finite number of players or agents pick from a
finite set of actions (possibly varying from agent to agent), at discrete
times, and receive rewards that depend on their own and the other agents'
actions.  Each agent chooses according to a rule called its {\em strategy}; if
agents have more than one strategy available, they use a meta-strategy to chose
among them.

As described, the setting fits classical game theory
\cite{von-Neumann-Morgenstern,Gintis-game-theory-evolving}, with its assumption
of so-called substantive rationality --- the agents have rational, accurate
expectations, enjoy common knowledge of the structure of the game, and act so
as to maximize their reward.  Casual observation suggests this is not a good
model of human behavior, and indeed experimental psychology
\cite{Kahneman-Slovic-Tversky} and experimental economics
\cite{Camerer-handbook} both reject it decisively.  Failing substantive
rationality, one turns to some form of {\em bounded rationality}
\cite{Simon-artificial} --- agents still pursue goals, but are limited by
finite knowledge and finite cognitive capacities, and so cannot, in general,
maximize anything.  Bounded rationality is the subject of a rich literature in
the social sciences, which encompasses both abstract models
\cite{Rubinstein-bounded}, some of which exploit familiar methods of
statistical physics \cite{Young-strategy-structure}, and a rapidly-growing
experimental literature, which includes its own array of models
\cite{Klein-sources-of-power,Zsambok-Klein-naturalistic,%
  Gigerenzer-Todd-heuristics,Gigerenzer-Selten-bounded}.  While the
econophysical literature almost exclusively considers boundedly-rational
agents, it never builds upon other work on bounded rationality.  For instance,
one of the best-supported meta-strategies is that of {\em satisficing}, in
which an agent searches over its options and takes the first one it finds with
a satisfactory reward
\cite{Simon-behavioral-model,Simon-rational-choice-and-environment,%
  Simon-models-of-bounded-1-and-2}.  We are unable to find a single paper which
employs this notion.  Econophysical strategies are all, so far as we know, of
the following form.  Each agent remembers a simple summary statistic (e.g., the
best action to have taken) for each of the last $m$ turns of the game.  For
each such {\em history}, a strategy specifies an action to take.  Rather than
maximizing, or even satisficing, they employ adaptive meta-strategies tending
to select strategies that did well in the recent past.

These caricatures or toy models have been studied since the 1980s, because, it
is said, they (1) shed light on real learning and mutual adaptation, and (2)
show surprisingly complex behavior.  In this paper, we examine two forms this
complexity is held to take --- deterministic chaos, which is a sort of
dynamical complexity, and formally measured complexity, specifically the amount
of information embedded in the system.  The rest of this section describes the
most popular discrete adaptive games, the minority game and the El Farol Bar
problem.  The next section introduces the tools from symbolic dynamics,
particularly subshifts of finite type, which we will use to pry open the
mechanisms of these games.  We show that the games are non-chaotic, and put
linear upper bounds on their complexities.  We discuss how to extend our
results to games with either memories of unlimited length, or noisy play.  We
then consider what you must do to a game to make it chaotic, if you want it to
be chaotic.  In closing, we suggest a way to reconcile our findings with common
observations on these games.

\subsection{Examples: El Farol and the Minority Game}

The best-loved game in econophysics is the {\em minority game} of Challet and
Zhang \cite{Challet-Zhang-minority-game}, the subject of a rapidly growing
literature (see \texttt{http://www.unifr.ch/econophysics/minority/}).  There
are $N$ agents, each of which can make only one of two moves: play ``0'' or
play ``1''.  If most agents play 0, then those who play 1 win, and vice-versa.
($N$ is always odd.)  Each player remembers whether 0 or 1 won on each of the
last $m$ turns.  Strategies, again, are rules mapping histories to actions;
each agent has $s$ of them.  In effect, strategies must predict what most
players will do, and then do the opposite.  Agents keep track of how many times
each of their strategies would have won, and use the strategy with the best
record over the time it can remember.  (Note that psychologists have studied
human performance in just such discrete prediction situations since the 1950s,
without finding anything resembling this learning mechanism
\cite{Myers-probability-learning}.)  Ties between strategies are broken
randomly or pseudo-randomly, say by assigning strategies a random order at the
beginning of the game.

Traditionally, the minority game has been studied either using the brute
empiricism of simulations, or with statistical-mechanical methods, particularly
those employing the thermodynamic limit.  Recently, Jefferies, Johnson et al.
have introduced methods owing more to dynamical systems, e.g., de Bruijn graphs
over possible sequences of winning outcomes
\cite{Jefferies-Hart-Johnson-deterministic,Jefferies-Lamper-Johnson-anatomy}.
Our approach here is related to theirs, but is even more deterministic, and
more microscopic.

The minority game descends from W. B. Arthur's El Farol Bar Problem
\cite{Arthur-increasing}.  There are 100 people in Santa Fe who go out to bars;
El Farol is the only good bar.  Normally, those who go there are better off
than those who stay home and watch the stars.  But El Farol is small, and if
more than 40 people go, crowding makes the bar-goers worse off than
stay-at-at-homes.  Everyone knows how many people went to the bar on each of
the last $m$ nights.  Strategies, again, map histories to actions --- staying
home or going to the bar.

Rather than statistical mechanics, the traditional approaches to the El Farol
problem are those of evolutionary adaptation and inductive behavior.  Many have
sought to show that either the population of agents co-evolves a set of
mutually-tolerable strategies, or that a ceaseless evolutionary arms race
forces everyone to ``run as fast as they can just to stay in place''
\cite{Arthur-increasing,Gintis-game-theory-evolving}.

\section{Symbolic Dynamics, Shifts}

Symbolic dynamics is one of the basic analytical methods of nonlinear dynamics
and complex systems, and as such its elements are discussed in standard
introductory textbooks on those subjects (e.g.,
Refs.\ \cite{Guckenheimer-Holmes,Devaney-first-course,Alligood-et-al-chaos,%
  Hilborn-chaos,Wiggins-baby,Badii-Politi}).  Strangely, however, it does not
seem to have been applied to discrete adaptive games before.  We therefore give
a few words of motivation, before presenting some basic notions and results.
Readers whose curiosities are aroused will find Refs.\ \cite{Wiggins-baby} and
\cite{Badii-Politi} particularly nice introductions, emphasizing uses in
nonlinear dynamics and complexity, respectively.  Refs.\ \cite{Kitchens} and
\cite{Hao-applied-symb-dyn} more advanced treatments.

\subsection{Why Symbolic Dynamics?}

The essence of symbolic dynamics is substituting strings of discrete characters
or symbols for trajectories in an original continuous state space.  Motion in
the original space corresponds simply to shifting along the string.  This idea
goes back to work by Morse and others in the 1930s.  One of its first great
successes came in a paper by Smale \cite{Smale-many-periodic-points} which
introduced his famous ``horseshoe'', and established the geometry of what we
now call chaotic dynamics.  Smale first constructed a differentiable,
invertible map $f$ on a two-dimensional manifold $M$ , which, while it has a
very simple functional form, has almost impossibly complicated geometric
behavior.  He then connected it to the full binary shift map $\sigma$ on
infinite sequences of 0s and 1s --- that is, the map which simply substitutes
the symbol at position $i+1$ for that at position $i$.  The dynamics of
$\sigma$ are very easy to grasp.  In particular, it is easy to see that it has
a countable infinity of periodic points of arbitrarily high period, an
uncountable infinity of non-periodic points, and a dense orbit.  (Recall that
rational numbers, which are countable, have repeating binary representations,
while irrational numbers, which are uncountable, never repeat.)  The great
trick, then, was to show that the horseshoe map $f$ and the shift map $\sigma$
were topologically conjugate.  That is, he found a continuous function $\phi$
from a subset of the manifold $M$ to the sequence space $\Sigma$ which was
one-to-one and onto, and which commuted with the two maps.  (Symbolically,
$\phi \circ f = \sigma \circ \phi$.)  This meant that the dynamical properties
of the horseshoe map could be read off directly from those of the shift map.

The attraction of symbolic dynamics should now be clear: we can, so to speak,
factor the dynamics of interest into a clean, simple shift map ($\sigma$), and
a messier, but less material, conjugacy map ($\phi$).  If, for instance, we
could show that the shift map and the minority game are topological conjugate,
or discern the mechanisms that prevent them from being conjugated, the dynamics
of the minority game itself would be much more plain to us.

\subsection{Basic Notions}

We now will quote the relevant standard results without proofs, which may be
found in e.g. Ref.\ \cite{Kitchens}.  Begin with a finite or countable {\em
  alphabet} $X$ of {\em symbols}.  From these we make {\em strings} or {\em
  words} over $X$, which are {\em symbol sequences}, $x_0 x_1 x_2 \ldots x_n$.
Our {\em shift space} is a set of semi-infinite words, $x = x_0 x_1 x_2 \ldots
x_i \ldots$ (i.e. we will be concerned with sequences on only one side of the
decimal point).  The dynamics are given by a {\em shift operator} $T$: $Tx =
x_1 x_2 \ldots$.  That is, $(Tx)_i = x_{i+1}$.  The {\em orbit} of a point is
its set of future iterates, $\left\{x,Tx, T^2x, \ldots\right\}$.

It is useful, and perhaps comforting, to invoke linear operators at this point.
Let $A$ be a square 0-1 matrix, with as many rows as $X$ has symbols.  Define a
shift space $X_A$ where $x$ is allowed iff $\forall i$, $A_{{(x_i)}{(x_{i+1})}}
= 1$.  Then $X_A$ is a {\em subshift of finite type} and the matrix $A$
corresponds to the shift operator $T$.  We can also represent $X_A$ with a
directed graph $G_A$, where there is one node for each symbol, and an edge from
$i$ to $j$ iff $A_{ij} = 1$.  Then $x \in X_A$ iff it is a path through $G_A$.
The transition matrix is {\em one-many} if there are symbols $i,j,k$ such that
$A_{ij} = 1$ and $A_{ik} = 1$ but $j \neq k$.  $A$ is {\em many-one} if there
are $i,j,k$ such that $A_{ik} = 1$ and $A_{jk} = 1$ but $i \neq j$.  And $A$ is
{\em one-one} if it is neither many-one nor one-many.

We now have a space $X_A$ of sequences, and some of the properties of this
space will be useful later on.  First, the standard metric between two
sequences (points) $x$ and $y$ in the shift space is
\begin{eqnarray}
d(x,y) & = & \sum_{i=0}^{\infty}{\frac{1-\delta(x_i,y_i)}{2^i}} ~,
\end{eqnarray}
where $\delta(a,b)$ is the Kronecker delta.  Thus if $x_i = y_i$ for all $i
\leq M$, then $d(x,y) \leq 2^{-M}$.  Conversely, if $d(x,y) < 2^{-M+1}$, then
$x_i = y_i$ for all $i \leq M$.  From this it follows that the sequence space,
equipped with the standard metric, is compact, totally disconnected and perfect
(i.e., it is closed, and every point is a limit point).

\subsection{Entropies and Lyapunov Exponents}

It is now useful to deal with notions defining growth rates of a given space
and spreading rates along orbits.  This will be achieved via standard entropies
and Lyapunov exponents borrowed from ergodic theory.  In general, not all words
are allowed, i.e., not every semi-infinite word is in the shift space.
Likewise, many of the sequences are never seen due to the particulars of the
shift operator.  If $W(L)$ is the number of allowed words of length $L$, the
{\em topological entropy} is
\begin{eqnarray}
\label{eqn:te-rate-defined}
h & = & \lim_{L\rightarrow\infty}{\frac{1}{L}\log{W(L)}} ~.
\end{eqnarray}
That is, $h$ is the asymptotic exponential rate of growth in the number of
allowed words.  It measures the degree of exponential spreading in the
dynamics.  Normally, of course, that is measured by the Lyapunov exponents.
Assuming those can be defined, the largest Lyapunov exponent $\lambda_1 \leq
h$.  (For specific conditions under which the Lyapunov exponents exist, see
Refs.\ \cite{Ruelle-exponents-in-Hilbert-space,Pesin-on-LEs,Katok-on-LEs}.)

The topological entropy gives equal weight to every {\em possible} word, no
matter how improbable.  If we wish to give more weight to the more probable
strings, it is natural to start with the entropy over words of length $L$,
\begin{eqnarray}
H(L) & \equiv & \left< - \log{\Prob(x^L_1)} \right>
\end{eqnarray}
where $\left< \right>$ denotes expectation.  This is the mean amount of
information, in bits, required to specify a word of length $L$ generated by the
dynamics.  We then define the {\em metric entropy rate} or {\em
  Kolmogorov-Sinai entropy} as
\begin{eqnarray}
\hmu & \equiv & \lim_{L\rightarrow\infty}{\frac{H[L]}{L}} ~.
\end{eqnarray}
In addition to being an important invariant of the dynamical system, the metric
entropy rate can be thought of as the rate of {\em information production} ---
the rate at which new information is produced by the dynamics.  Clearly, $h
\geq \hmu$ \cite{Katok-Hasselblatt}.  Under certain conditions on $f$
\cite{Ruelle-inequality,Margulis-negative-curvature,Pesin-on-LEs}, $\hmu$ is
equal to the sum of the positive Lyapunov exponents; more generally it is a
lower bound on their sum.  Thus, a positive $\hmu$ guarantees the existence of
at least one positive Lyapunov exponent.  The proofs of these assertions are
somewhat complicated results from ergodic theory, well-summarized in
Ref.\ \cite{Katok-Hasselblatt}.

\subsection{Chaos}

We adopt the standard topological definition of chaos, due to Devaney
\cite{Banks-et-al-on-Devaney}.  A compact set $\Lambda$, invariant under $T$,
is {\em chaotic} if $T$ is topological transitive on $\Lambda$, and its
periodic orbits are dense in $\Lambda$.  {\em Topological transitivity} means
that the trajectory of any open set in $\Lambda$ eventually visits every other
open set in $\Lambda$ --- we can get from one part of the region to any other.
Together with the requirement of dense periodic orbits, this implies sensitive
dependence on initial conditions.  One can imagine the chaotic trajectories
wandering between (but never on) the dense periodic orbits.  Note that the
periodic orbits, while dense, are not stable or attracting \cite{Smale-dds}.
In fact, sensitive dependence on initial conditions, which we have just
established, in turn implies exponential separation of nearby orbits, at least
under fairly weak conditions, and hence a positive Lyapunov exponent and
positive entropy rates.  These, accordingly, constitute the numerical or
experimental signatures of chaos.

For subshifts of finite type, all of this can be related to the matrix $A$ (or
the corresponding graph) very directly.  $X_A$ is {\em transitive} if it has
dense orbits.  (N.B., not necessarily dense {\em periodic} orbits.)  $X_A$ is
{\em irreducible} if for every pair of indices $i$ and $j$, there is some $l>0$
for which $\{ A^l \}_{ij} > 0$.  (This is the same as $G_A$ being strongly
connected.)  Now, if $X_A$ is irreducible, then either it is transitive or it
consists of a single periodic orbit.  Likewise, if $X_A$ is transitive, then
$A$ is irreducible.

\section{The Minority Game as a Shift Map}

To do one iteration of the minority game, we need to know two things.  (1) The
history of the last $m$ rounds --- which action won on each of them.  (2) The
``gains'' of all the strategies possessed by the agents --- the number of times
each strategy predicted the winning action in the last $m$ rounds.  These facts
fix the action of each agent, and so the winner of the round.  That is, the
next state of the game is a deterministic function of its present state.  (In
what follows, ``state'' will always mean the global state of the game, never
the state of an individual agent.)

Since each state variable takes only finitely many values, we can map the state
as a whole into a finite alphabet.  We order the states and their alphabetic
symbols however we like.  Then we build a square matrix $A$, where $A_{ij} = 1$
if $j$ follows $i$, and $=0$ otherwise.  $X_A$ defines a subshift of finite
type over states of the minority game, and every run of the minority game is an
allowed orbit of this subshift.  We could write down $A$ in some detail, but
shan't.

Suppose memory for gains extends back indefinitely.  Then the state comes from
a countably infinite alphabet, rather than a finite one, but the subshift is
still deterministic (many-one).

In the finite-memory case, the lack of chaos is elementary once we realize that
there are only finitely many states; the full apparatus of symbolic dynamics is
over-kill.  There are only finitely many symbols, so every string will
eventually repeat a symbol.  But each symbol has a unique successor, so after
the first repetition, the string just repeats a single word.  Hence every orbit
is eventually periodic.

The topological entropy rate is zero.  The first symbol of a string fixes all
subsequent ones, so $\forall L$, $W(L) = W(1)$, and $h=0$ by
Eq.\ \ref{eqn:te-rate-defined}.

\section{Complexity in the Minority Game}

Before we can say quantitative things about the complexity of a model, we must
pick a complexity measure, of which there are many \cite{Badii-Politi}.  When
people have discussed the complexity of discrete games at all, they have picked
the Kolmogorov complexity or ``algorithmic information content''
\cite{Li-and-Vitanyi-1993}.  Kolmogorov introduced this concept as part of a
definition of ``randomness'' relying on purely combinatorial notions, part of a
long-standing program of research in the foundations of probability
\cite{von-Plato-modern-prob}.  The Kolmogorov complexity $K(w)$ of a finite
string $w$ is the length of the shortest computer program which will produce
that string as its output and then stop\footnote{We will ignore some subtleties
  relating to the choice of programming language, which do not affect the
  argument}.  Because any string can be produced by a program like
\texttt{print ``w''}, the complexity of any string cannot be more than its
length plus a constant (the length of the ``print'' instruction).  On the other
hand, some strings can be produced by programs which are much shorter than
themselves, e.g., a string of a million 0s can be produced by iterating the
command \texttt{print ``0''} a million times.  A random strings, roughly
speaking, is one which cannot be generated by any program significantly shorter
than the string itself.  More exactly, if $x^L_1$ are the first $L$ symbols of
an infinite word, and $K(x^L_1)/n \rightarrow 1$, then all the properties which
are true of a sequence of independent, identically distributed random variables
are true of the symbols in the infinite word.  For deep reasons connected with
the existence of uncomputable numbers, there can be no general procedure for
computing the Kolmogorov complexity of arbitrary strings, nor are there any
known procedures which give well-controlled approximations
\cite{Li-and-Vitanyi-1993,Cover-and-Thomas}.  However, if the symbols are
generated by an ergodic process, there is an easy expression for the expected
rate of growth in the complexity \cite{Brudno-1978}: $\left< K(x^L_1) \right>
/L \rightarrow \hmu$ as $L \rightarrow \infty$.  This is another illustration
of the fact that $\hmu$ measures the rate of information production.

While, as we have said, there is no real way to compute the Kolmogorov
complexity, several papers in the physics literature have attempted to
calculate it for games
\cite{Mansilla-alg-compl-in-MG-in-compl-sys,Mansilla-alg-compl-in-MG-in-PRE,%
  Mansilla-from-naive-to-sophisticated} and even for real markets
\cite{Mansilla-complexity-of-real-markets,Mantegna-Stanley-econophysics}, with
the claim \cite{Mantegna-Stanley-econophysics} that the high complexity they
obtain shows that financial time series contain much valuable information.
{\em Pace} these authors, had they actually been able to calculate the
algorithmic information content, a high value of it would indicate that the
systems they considered were basically random, because this is just what
Kolmogorov complexity {\em means}.

Assuming one wants a notion of ``complex'' which is not the same as ``random'',
what should a good complexity measure be like?  As mentioned, many measures
have been proposed \cite{Badii-Politi}, but there is some consensus about what
features a good measure should have.  It should be low for both highly ordered
and highly random systems; it should be calculable; and it should have a clear
dynamical meaning, rather than just being a number for its own sake
\cite{DPF-JPC-why}.  To our knowledge, only one measure satisfies all these
criteria, the {\em statistical complexity} of Crutchfield and Young
\cite{Inferring-stat-compl}, a fully formalized and operational version of
Grassberger's \cite{Grassberger-1986} conjectural ``true measure complexity''.
It is the amount of information about the past of the system needed for
maximally accurate prediction of its future \cite{CMPPSS}.

We get the statistical complexity by finding the {\em causal states} of the
system: each causal state is a distinct distribution for future events,
conditional on the past of the system.  Abstractly, we find the conditional
distribution $\Prob(x^\infty_1|x^0_{-\infty})$ for each past $x^0_{-\infty}$,
and group pasts with the same distribution into states.  The probability of a
state is then the sum of the probabilities of all the pasts assigned to it, and
the distribution conditional on the state is just the same as the distribution
conditional on the individual pasts.  (In practice, we use a number of
recursive tricks to approximate the causal states from finite data, while only
looking at the conditional distribution of the next symbol \cite{AfPDiTS}.)
The statistical complexity $\Cmu$ is simply the entropy (uncertainty in bits)
of the causal states $= -\sum_{i}{p_i\log_2{p_i}}$, where $p_i$ is the
probability of the $i^{\mathrm{th}}$ causal state.  This is maximized when all
causal states are equally probable.  In the case of a periodic process, each
phase is a causal state, and each state is equally probable, so $\Cmu =
\log_2{P}$.\footnote{For non-periodic processes, $2^{\Cmu}$ is the geometric
  average of the expected recurrence times of the causal states, i.e., still
  the average amount of time needed to return to a state after leaving it.}

We have just seen that the minority game has only periodic orbits, so to get
$\Cmu$, we just need the period.  Start with just one strategy per agent.
After the system has seen every possible history, it must repeat itself, so $P
\leq 2^m$.  Now give each agent several strategies.  The number of ways agents
can deploy strategies is $S \leq s^N$.  When we've gone through every possible
history, the agents could be using different strategies than before.  But once
we've seen all deployments of strategies, we must be back where we started.
Thus $P \leq S 2^m \leq s^N 2^m$.  Even these bad estimates give us
\begin{eqnarray}
C_{\mu} & \leq & m + N\log_2{s}
\end{eqnarray}
The complexity grows linearly with memory and population \textit{at most}.  We
get this growth rate only with complete (combinatorial) independence --- an
agent's strategy selection is constrained neither by the selections of other
agents, nor even by the global history!

In a number of recent papers, Savit, Parunak and collaborators
\cite{Savit-and-Altarum-general-structure,Savit-and-Altarum-phase-structure,%
  Parunak-et-al-effort-profiles} have considered a complexity measure closely
related to the statistical complexity (they do not give it a name).  In
essence, following \cite{Savit-dependent-variables}, they treat two pasts as
belong to the same state if and only if they they lead to the same {\em unique}
next symbol, whereas statistical complexity requires only that they lead to the
same {\em distribution} of future symbols.  The two complexities coincide for
periodic processes.  In the presence of noise, however, the method of Savit et
al.\ will be forced to rely on longer and longer histories, until it finds ones
that look deterministic simply because of undersampling, and their complexity
will diverge logarithmically in the size of their data set
\cite{Marton-Shields}.  The statistical complexity, however, is well-behaved in
the presence of noise (see Section \ref{sec:noise} below).

\section{Infinite Memory for Gains}

People sometimes play the minority game with an infinite memory for gains ---
agents remember the number of times each strategy has won since the beginning
of the game, rather than just in the last $m$ turns. Then the alphabet is
countably infinite.  But what matters are the difference in strategies' gains,
and if those are bounded, the finite-state arguments take over.  Even when
differences in gains are unbounded, the shift map is many-one, so there is just
a single 1 in each row of the transition matrix $T$.

The topological entropy rate of a countable-state shift is \cite[Observation
7.2.10]{Kitchens}
\begin{eqnarray}
\label{countable-TE-formula}
h & = & \limsup_{n\rightarrow\infty}{\frac{1}{n}\sum_j{(T^n)_{ij}}} ~,
\end{eqnarray}
independent of $i$.  From many-oneness, the sum in Eq.\
\ref{countable-TE-formula} is always exactly 1, so $h = 0$.

Let us consider also the question of exponential spreading.  

The distance between points in the standard metric is at most 2.  $d(x,y) = 2$
iff $x_i \neq y_i$ for all $i$.  In that case, $d(T^n x,T^n y) = 2$ for all
$n$, so maximally-separated points stay maximally separated.  On the other
hand, if the distance between $x$ and $y$ is less than 2, then at some
position, $k$, $x_k = y_k$.  By the many-oneness of the shift, if $j > k$, $x_j
= y_j$.  Therefore, the distance between the iterates of $x$ and $y$ shrinks
until it hits 0.  Distance never increases; spreading of any sort, never mind
exponential spreading, is impossible.

The distance between any two distinct periodic points is always maximal; that
is, given two periodic points $x$ and $y$, if $x \neq y$, then $x_i \neq y_i$
for all $i$.  To see this, consider first the case of two points on the same
periodic cycle, of period $P$.  For some $0 < L < P$, $T^L(x) = y$.  Suppose
that, for some $k$, $x_k = y_k$.  Then $T^k x = T^k y = T^{k+L} x$, so $T^k x$
is periodic with period $L < P$, which is absurd.  Now consider points on two
distinct periodic cycles.  Suppose there were a $k$ such that $x_k = y_k$.
Then $T^k(x) = T^k(y)$.  Hence there is a point, $T^k(x)$, belonging both to
the $x$ cycle and the $y$ cycle.  But this is absurd, because the cycles are
distinct.

Suppose periodic points were dense.  Then for any point $x$ and any distance
$\epsilon > 0$, there would be a periodic point $y \neq x$ such that $d(x,y)
\leq \epsilon$.  But if $x$ is itself periodic, then we have just seen that
$d(x,y) = 2$ no matter what periodic point $y$ we chose.  Therefore periodic
points are not dense.  Chaotic maps have dense periodic points, so this shift
map isn't one.

\section{Noise}
\label{sec:noise}

Traditionally, the minority game is played somewhat noisily --- either players
chose strategies with some degree of randomness, or the move they make, once
they have chosen a strategy, is somewhat random.  This makes the game amenable
to statistical-mechanical treatment, a point which has been thoroughly explored
elsewhere.  Here, we will see that it can make the entropy rate positive, but
does not increase the complexity.

The usual types of noise imposed in discrete games change them from
deterministic maps into Markov chains.  This means we can represent the state
of the game at one time as a function of the state at the previous time, and a
sequence of stationary, independent noise variables, i.e., $x_{t+1} =
f(x_{t},\eta_t)$, where $\left\{\eta_t\right\}$ are IID random variables which
are also independent of the $x_t$.  The Markov property of the sequence
$\left\{x_t\right\}$ also means that its entropy rate (metric or topological)
is just the entropy of $x_{t+1}$ conditional on $x_t$
\cite{Cover-and-Thomas,Kitchens}.  That is, $\hmu = H[x_{t+1}|x_{t}]$, and
similarly for $h$.  Now, applying some basic results of information theory
\cite[ch.\ 2]{Cover-and-Thomas},
\begin{eqnarray*}
H[x_{t+1}|x_t] & = & H[f(x_t,\eta_t)|x_t] \\
& \leq & H[x_t,\eta_t|x_t]\\
& = & H[\eta_t|x_t] = H[\eta_t]
\end{eqnarray*}
Or, in words: because the only source of randomness is the noise, the entropy
rate of the states must be no more than the entropy of the noise.  A parallel
argument holds for the topological entropy.  Nothing in the process
{\em amplifies} the noise.

Now let us consider the statistical complexity.  Recall that a causal state is
a distinct distribution of the system's future behavior, conditional on its
past.  For noise of the sort described, the present state of the game
determines the \textit{distribution} of future symbols, independent of any
other feature of the past of the system.  So the present symbol fixes the
causal state.  The noise could make some causal states more likely than others
(reducing the complexity), or could even lead to two symbols having the same
distribution over future orbits, merging them into a single causal state (again
lowering the complexity).  Adding noise simplifies matters, if it does anything
at all.

\section{Conclusions}

We analyzed the global behavior of discrete adaptive games, using symbolic
dynamics.  These games are not chaotic, and the most popular ones are actually
of low complexity, at most linear in the number of agents.

The crucial obstacles to chaos were that the state of the minority game can be
summarized in a discrete alphabet, and that the evolution of those states is
many-one.  Our results will apply, \textit{mutatis mutandis}, to any
deterministic adaptive game of which this is true, for instance, the El Farol
problem, and various discrete market models.  To repeat, this is not an
artifact of discretizing continuous models, but a feature of deliberately
discrete ones.  Nothing depends on the meta-strategy; our results hold with
learning, imitation, evolution, etc., so long one implements them
non-anticipatively.

\subsection{Exponential Spreading  versus Exponential Transients}

Anyone who looks at a simulation of the minority game sees an intricate,
hard-to-predict process --- we certainly do.  Scientists trained in nonlinear
dynamics naturally look to exponential divergence of orbits to explain these
observations.  We have seen that there is no such divergence in the standard,
noiseless version of these games; there are three ways to get it.
\begin{itemize}
\item \textit{Hidden information.}  If only some agents have access to a
  (changing) outside source of information, then there can be apparent
  transitivity in the game, and so deterministic chaos.  A neat example is the
  Brock-Hommes asset-trading model
  \cite{Brock-Hommes-heterogeneous-beliefs,Brock-Hommes-rational-randomness,%
    Brock-Hommes-rational-animal-spirits}, where agents get simple adaptive
  strategies for free, but must pay for rational expectations of future prices.
\item \textit{Add noise.}  As we've seen (Sec.\ \ref{sec:noise}), this
  certainly allows exponential divergence, but in a trivial way, and at a cost
  in complexity.
\item \textit{Coupling to chaos.}  Break ties using a chaotic shift map, or
  something similar.  Often this will make the game chaotic.
\end{itemize}
Only the first of these mechanisms has any {\em economic} relevance.  In real
markets, information is expensive, but transparency is prohibitively expensive,
and so hidden information is ubiquitous.  Exploring the consequences of this is
an important problem of great economic interest; indeed, the problem is common
across almost all social situations.  The other methods, by contrast, are
``unphysical,'' lacking in substantive content.  Of course, hidden information
cannot explain the phenomenology of games in which it is absent, but it is no
less interesting on that account.

What however {\em does} account for that phenomenology?  We suggest that the
answer is exponentially fast divergence, but rather exponentially slow
convergence.  For the minority game, it is numerically well established that
relaxation times grow exponentially with $m$
\cite{Marsili-Challet-continuum-time-limit}.  What is less appreciated is that,
in high-dimensional dynamical systems, the approach to the attractor can be so
slow that it is never reached in simulations, at least not within the lifespan
of ordinary experimenters (but see \cite{Dyson-time-without-end}).  In systems
formed by coupling smaller dynamical systems, such as the agents in our games,
the duration of the transient epoch often grows exponentially with the number
of agents \cite{JPC-Kaneko-attractors}.  The attractors and their invariant
distributions are irrelevant; what matters is the structure of the attractor
basin \cite{Wuensche-Lesser} and the quasi-stationary transient distribution it
produces \cite{JPC-subbasins}.  Once recognized, these long transients are
actually a {\em feature} of discrete adaptive games, since the social systems
they are supposed to model are essentially never in an invariant state, but
always ``in transition''\footnote{This does not mean they are never in
  equilibrium, in the economists' sense, however.}.  The transients are,
however, bad news for methods based on either ergodic or thermodynamic limits.
Quantities like the entropy rate and the Lyapunov exponents, which are
rigorously defined only in the limit, loose much of their interest.

\subsection{Directions for Future Work}

The motive for studying models like the minority game should {\em not} be that
they are complex systems in some formal sense.  In any area of modeling, formal
constructs acquire interest from their ability to help resolve problems of
physical relevance --- that is, here, social relevance.  We have argued that
such constructs as algorithmic complexities and stationary and thermodynamic
limits have not demonstrated such relevance {\em here}, and are unlikely to do
so in the future.  By contrast, the distinctively transient dynamics of such
systems, especially their basin structure, hold great promise for understanding
the origins of complex behavior we see in these games.  The hope of
econophysics is that these models portray, in stylized miniature, key aspects
of real mutual adaptation and social interaction, and that studying the
portrait will help us understand the original.  This is not an outlandish hope,
since profound insights have come from studying the equally-stylized Prisoners'
Dilemma \cite{Axelrod-evol-of-coop}, and a host of other, less famous models
\cite{Simon-artificial,Gintis-game-theory-evolving,Schelling-micro-macro} have
illuminated the collective life of political, boundedly-rational animals.  We
hope that the discrete adaptive games of econophysics will take their place in
this set.  The contribution of econophysics, it seems to us, be to assimilate
the rich body of empirical findings and models of individual choice and
learning, and say something distinctive and illuminating about how large
numbers of such boundedly-rational agents interact.

\subsection*{Acknowledgments}

We thank Sam Bowles, Sven Brueckner, Jim Crutchfield, Dee Dechert, Richard
Gonzalez, Cars Hommes, Matt Jones, Cris Moore, Scott Page, Van Parunak, Robert
Savit, Kris Shalizi, Clint Sprott and Libby Wood for valuable conversation and
suggestions.  CRS also thanks Kris Shalizi for being perfect.  Our work at SFI
was supported by the Dynamics of Learning program, under DARPA cooperative
agreement F30602-00-2-0583, and CRS's work at Ann Arbor by a grant from the
James S. McDonnell Foundation.

\bibliographystyle{apsrev}
\bibliography{locusts}
\end{document}